\documentclass[cits]{PoS}
\pdfoutput=1
\usepackage{graphicx}
\usepackage{amssymb}
\usepackage{fontenc}
\usepackage{times}
\usepackage{mathptmx}
\newcommand{\Msun}{\mathrm{M}_{\odot}}
\newcommand{\Rsun}{\mathrm{R}_{\odot}}
\newcommand{\Rd}{R_{\mathrm{d}}}
\newcommand{\RL}{R_{\mathrm{L,1}}}
\newcommand{\Fe}{\mathrm{Fe}}
\newcommand{\Hy}{\mathrm{H}}
\newcommand{\C}{\mathrm{C}}

\newcommand{\Ba}{\mathrm{Ba}}

\title{The elusive origin of \\Carbon-Enhanced Metal-Poor stars}

\ShortTitle{The elusive origin of CEMP stars}

\author{\speaker{Carlo Abate}\\%
        Department of Astrophysics/IMAPP, Radboud Universiteit Nijmegen, Nijmegen, The Netherlands\\
        E-mail: \email{C.Abate@astro.ru.nl}}

\author{Onno R. Pols\\
        Department of Astrophysics/IMAPP, Radboud Universiteit Nijmegen, Nijmegen, The Netherlands\\
        E-mail: \email{O.Pols@astro.ru.nl}}

\author{Robert G. Izzard\\
        Argelander Institut f\"ur Astronomie, Universit\"at Bonn, Bonn, Germany\\
        E-mail: \email{izzard@astro.uni-bonn.de}}

\author{Shazrene S. Mohamed\\
        South African Astronomical Observatory, Observatory, South Africa\\
        E-mail: \email{mohamed@saao.ac.za}}

\author{Selma E. de Mink\thanks{Hubble fellow}\\
        Space Telescope Science Institute, Baltimore, MD, USA\\
        E-mail: \email{demink@stsci.edu}}

  \abstract
   {Very metal-poor stars of the halo are the relics of the star formation in the early Galaxy. 
   Among these stars carbon-enhanced metal-poor (CEMP) stars count for 9--25\%. 
   Most CEMP stars are also enriched in $s$-process elements and are called CEMP-s
   stars. The large number of CEMP-s stars observed in binary systems suggests that the chemical enrichment
   is due to wind mass transfer in the past from an AGB donor star on to a low-mass companion,
   the star that we are observing today.
   However, binary population synthesis models predict CEMP fractions of only $\approx2\%$. As an alternative
   to the canonical Bondi-Hoyle-Lyttleton (BHL) wind accretion model, recent hydrodynamical simulations
   suggest an efficient mode of wind mass transfer, called wind Roche-lobe overflow (WRLOF), can reproduce
   observations of AGB winds in binary systems. We use our population synthesis model to
   test the consequences of WRLOF on a population of CEMP stars. Compared to previous predictions based
   on the BHL model we find a modest increase of the fraction of CEMP stars and substantial differences in
   the distributions of carbon and periods in the population of CEMP stars.}
 
\FullConference{XII International Symposium on Nuclei in the Cosmos,\\
		August 5-12, 2012\\
		Cairns, Australia}

\begin{document}

\section{Carbon-Enhanced Metal-Poor stars}

The stars that we observe in the Galactic halo are typically low in mass, poor in metals and have hardly evolved since their formation.
We call very metal-poor (VMP) those stars characterised by an iron abundance $[\Fe/\Hy]$\footnote{given two elements X and Y, their abundance ratio is [X/Y]$= \log_{10} (N_X/N_Y) - \log_{10} (N_{X\odot}/N_{Y\odot})$, where $N_{X,Y}$ refers to the number density of the elements X and Y and $\odot$ denotes the abundance in the Sun.} $\lesssim-2.0\,$. Among the VMP stars of the halo the wide field spectroscopic surveys HK \cite{BeersAJ92} and HES \cite{ChristliebAA01} observed a large fraction of carbon-enhanced metal-poor stars ($[\mathrm{C}/\mathrm{Fe}] \ge +1.0$; CEMP stars). The observed CEMP to VMP ratio varies from $9\%$ to $25\%$ \cite{Frebel2006, Lucatello2006, Marsteller2005} and increases rapidly for decreasing iron content and increasing distance from the Galactic plane \cite{Carollo2012}. Most observed CEMP stars show an enhancement of heavy elements produced by slow neutron-capture process, e.g. barium. CEMP stars with barium abundance $[\Ba/\Fe]>+0.5$ are called CEMP-$s$ stars and account for approximately $80\%$ of all CEMP stars \cite{Aoki2007}.
 
Three different solutions have been proposed to explain the formation of CEMP stars: ($i$) CEMP stars are formed in binary systems where, in the past, the carbon-rich wind of a thermally-pulsing asymptotic giant branch (TPAGB) star was accreted to the surface of the companion which is the star that we are observing today; ($ii$) CEMP stars produce their own carbon in their interior and bring it to the surface by mixing episodes that happen in some stellar evolution models at extremely low metallicity \cite{Fujimoto2000}; ($iii$) the carbon is primordial and was produced by the first generation of stars \cite{Mackey2003}. The binary scenario is the most likely formation mechanism for CEMP-s stars: a spectroscopic analysis shows that the fraction of CEMP-s stars with detected radial-velocity variations is consistent with the hypothesis of all being members of binary systems \cite{Lucatello05b}.

Several studies attempt to reproduce the observed CEMP fraction of $9-25\%$ but the typical predictions of population synthesis models are at least a factor four lower than the observations \cite{Izzard09}. Some authors suggest that an initial mass function (IMF) biased towards intermediate-mass stars at low metallicity is required to reproduce the CEMP/VMP fraction measured in the halo \cite{Lucatello05a, Komiya07}, but large changes to the IMF are inconsistent with the small observed fraction of nitrogen-enriched metal-poor stars \cite{Pols2012}. The IMF at low metallicity is not the only uncertainty that affects the study of the origin of CEMP stars; other physical parameters are poorly constrained, e.g. the internal mixing and diffusion processes in stars, the nucleosynthesis during the AGB phase, the mass-transfer process, the binary fraction of low-mass stars in the halo and the initial distribution of the orbital parameters. For a description of these uncerainties in the context of CEMP stars we refer to the work of Izzard and collaborators and references therein \cite{Izzard09}. In this work we investigate the wind mass-transfer process and its effects on a population of CEMP stars.

\section{Wind mass transfer in binary stars}
In the late stage of its evolution a star of mass between $0.8-8\,\Msun$ becomes a TPAGB star. In this phase a sufficiently massive star produces carbon and $s$-elements in its interior, brings them to the surface by third dredge up and expels them in the strong wind mass loss characteristic of TPAGB stars. If the wind-losing star belongs to a binary system, a fraction of the ejected material is accreted to the companion star and pollutes its surface. The TPAGB star then evolves into a white dwarf and today we are able to observe only the chemically-enriched secondary star.
In this picture the wind mass transfer therefore has a fundamental role for the formation of CEMP stars.

The canonical description of wind mass transfer is the Bondi-Hoyle-Lyttleton (BHL) prescription \cite{BoHo}, which predicts low accretion efficiencies. The BHL prescription is based on the assumption that the wind velocity is much larger than the orbital velocity of the accreting star. However this assumption is not always valid in the case of AGB winds. Outflows from AGB stars are observed with velocities in the range $5-30$ $\mathrm{km\,s}^{-1}$ while the orbital velocity in binary stars of periods around $10^4$ days is about $10\,\mathrm{km\,s}^{-1}$, i.e. comparable. 

\begin{figure}
   \includegraphics[width=0.45\textwidth]{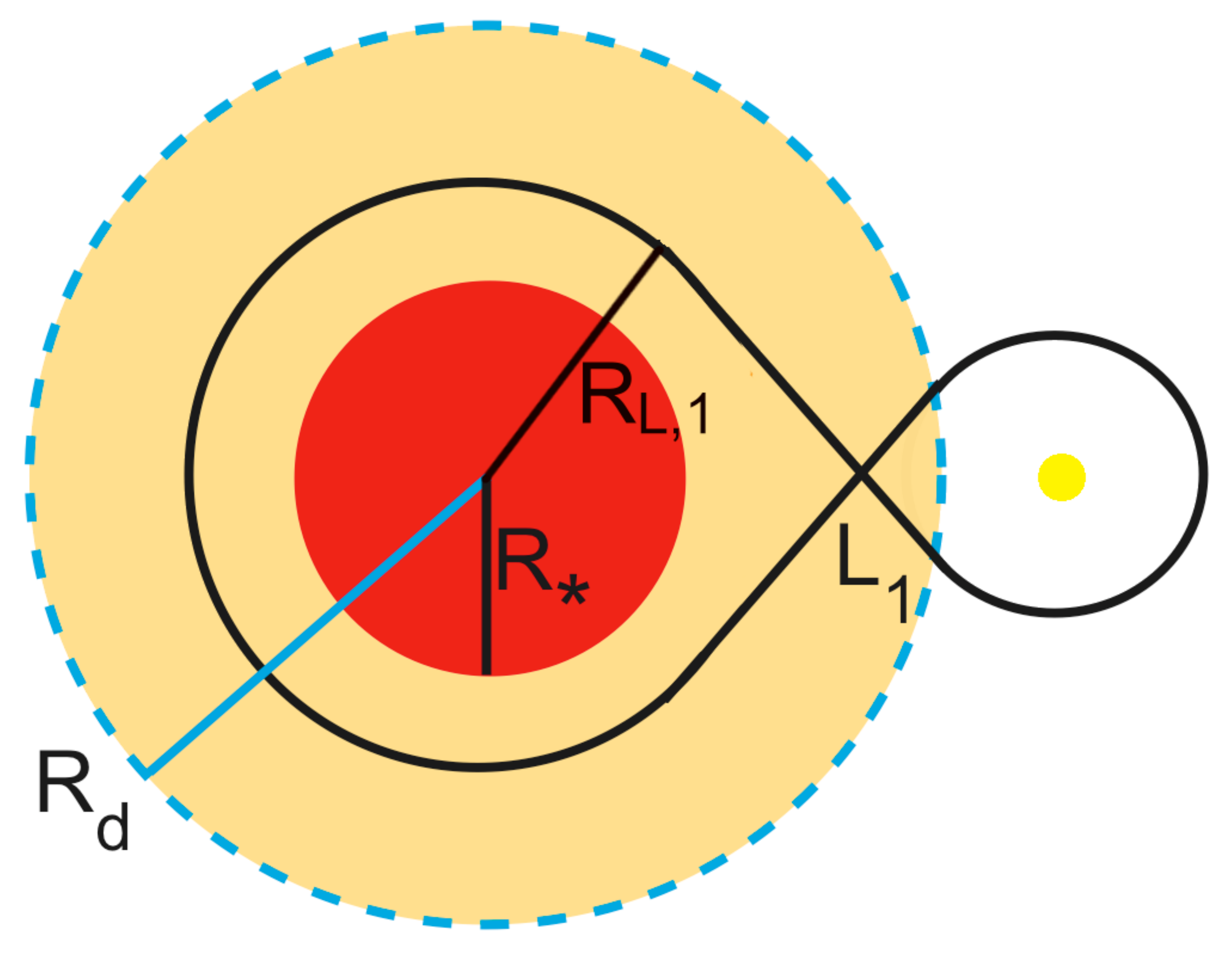}
   \includegraphics[width=0.45\textwidth]{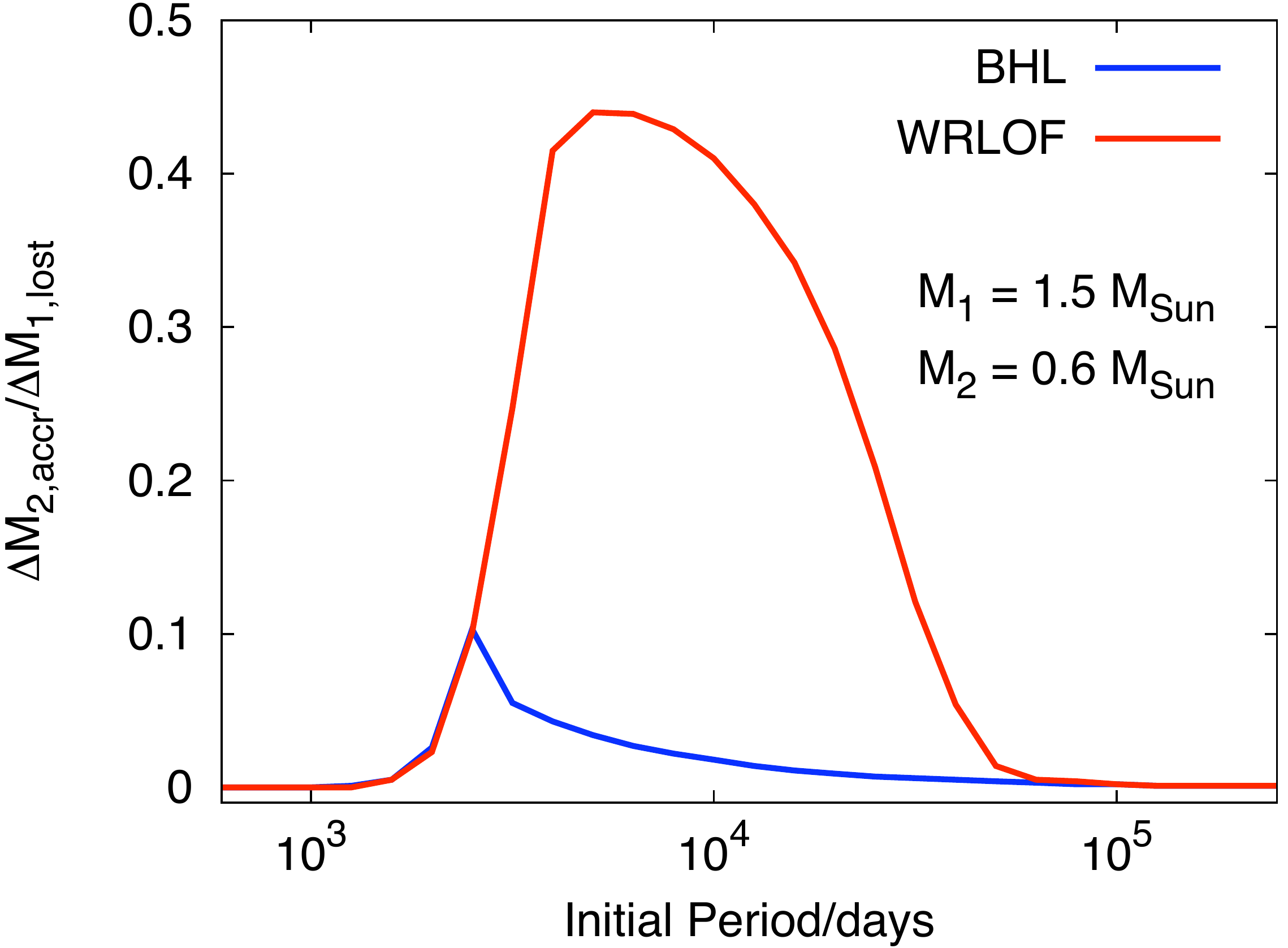}
      \caption{{\it Left panel.} Schematic picture of the WRLOF mechanism: the wind acceleration radius $\Rd$ lies close to the Roche-lobe radius $\RL$ of the AGB star (in red). The wind is slow inside the wind acceleration zone (orange area) and can be efficiently accreted to the low-mass secondary through the inner Lagrangian point $L_1$ (sizes are not in scale). {\it Right panel}. Efficiency of wind mass transfer $\beta_{\mathrm{acc}}$ as a function of the initial period for a binary system with $M_1=1.5\,\Msun$ and $M_2=0.6\,\Msun\,$. In this and in the plots that follow we show models calculated with BHL and WRLOF prescriptions in blue and red respectively.}
   \label{fig:WRLOF}
   \end{figure}

Recent hydrodynamical simulations suggest a mode of mass transfer called ``wind Roche-lobe overflow'' (WRLOF) applicable in the case of the slow AGB winds. WRLOF occurs in systems where the radius of the wind acceleration zone, i.e. the region where the wind is accelerated beyond the escape velocity, is larger than, or is a significant fraction of, the Roche-lobe radius of the wind-losing star. When this condition is fulfilled the wind is gravitationally confined to the Roche lobe of the donor star and is funnelled through the inner Lagrangian point $L_1$ towards the secondary, on to which it accretes with high efficiency (up to $50\%$) over a wide range of separations \cite{Shazrene07}. We schematically represent the conditions in which the WRLOF mode of mass transfer occurs in the left panel of Fig. \ref{fig:WRLOF}.

AGB winds are thought to be accelerated by a combination of stellar pulsation and radiation pressure. The pulsations increase the gas density in the cold outer atmosphere where dust grains form. The dust grains are accelerated by the radiation pressure and drag the surrounding gas along due to collisional momentum transfer \cite{Hofner2012}. 
Therefore, we assume that the wind acceleration radius coincides with the dust formation radius $\Rd$, which depends on the stellar radius, on the effective temperature of the star and on the condensation temperature of the dust (the higher this temperature is, the closer to the stellar surface the dust forms). 
During most of the AGB phase the dust formation radius is linearly proportional to the stellar radius and for carbon-rich dust $\Rd/R_*\approx3$.

In this work we implement in our binary population synthesis code a model of the WRLOF process based on the results of the above-mentioned hydrodynamical simulations \cite{Abate2012}. We present the results of the comparison between the evolution of a population of binary systems in the case of wind mass transfer modelled by the BHL prescription and by WRLOF.

\section{Results}

Our population synthesis simulations consist of $N^3$ binary stars in $\ln M_1 - \ln M_2 - \ln a $ parameter space, where $M_{1,\,2}$ are the initial masses of the primary and the secondary star, $a$ is the initial separation and we choose $N=128$. The initial masses $M_1\,$ and $M_2$ vary respectively in the range $[0.8,\,8.0]\,\Msun\,$ and $[0.1,\,0.9] \Msun$. The initial separation is chosen in the range $[3,\,10^5]\, \Rsun$ and we consider a binary fraction of unity in this range. The metallicity is $Z=10^{-4}$ which corresponds to $[\Fe/\Hy]\approx-2.3$. Stars are selected from our model population according to their age, surface gravity and surface abundances: VMP stars are older than 10 Gyr and with surface gravity $\log_{10}(g/$cm\,s$^{-2})\le 4.0$\,, i.e. we select essentially subgiants or giants; CEMP stars are VMP stars with $[\C/\Fe]\ge+1.0$. For a complete discussion of our model, its characteristics, parameters and uncertainties we refer to the work of Izzard and collaborators and our forthcoming paper \cite{Izzard09, Abate2012}.

In the right panel of Fig. \ref{fig:WRLOF} we show an example of the effect of our model of WRLOF on the accretion efficiency of wind mass transfer in a binary system with $M_1= 1.5\,\Msun\,,~ M_2=0.6\,\Msun$ and initial period in the range $[10^3\,,10^5]$ days. 
We compare the accretion efficiency $\beta_{\mathrm{acc}}$ calculated with the BHL prescription (blue line) and with our WRLOF model (red line). The maximum accretion efficiency in the case of WRLOF is more than $40\%$ and a high efficiency is possible over a wide range of periods. A $1.5\,\Msun$ star loses approximately $0.8\,\Msun$ during the AGB phase, therefore its $0.6\,\Msun$ companion can accrete enough mass to become carbon-enhanced. The maximum accretion efficiency in the BHL model is about $10\%$ for periods around $3000$ days, which means the companion accretes $0.08\,\Msun$ or less. A $0.68\,\Msun$ star is not massive enough to evolve off the main sequence and is not selected as a CEMP star by our model.

\begin{figure}
   \includegraphics[width=0.45\textwidth]{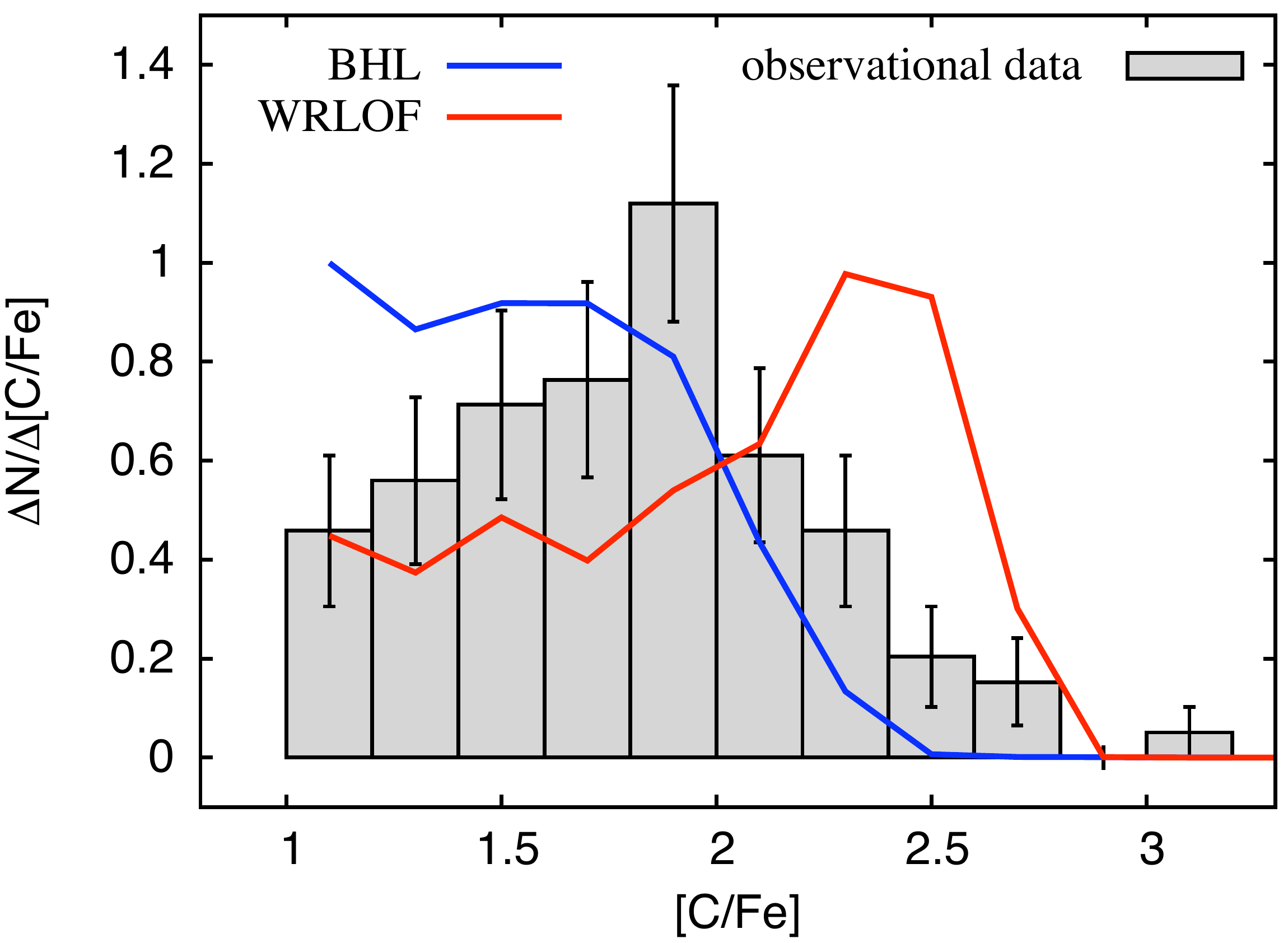}
   \includegraphics[width=0.45\textwidth]{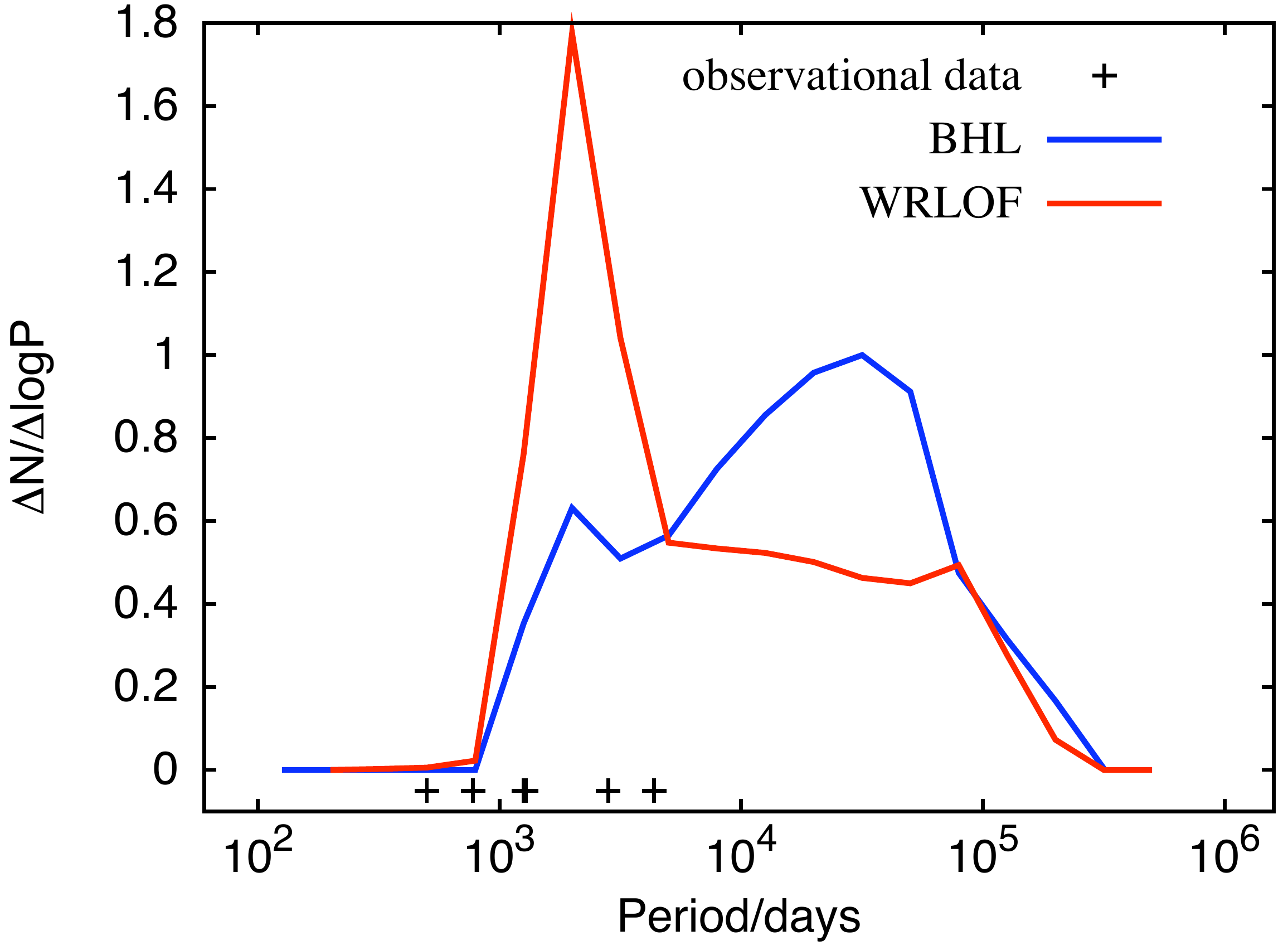}
      \caption{Distribution of [C/Fe] ({\it left panel}) and periods ({\it right panel}) in the CEMP population. Histograms show the observed distribution in our data sample, with Poisson errors. Plus signs are periods of observed CEMP stars. The bins are equally spaced and the width of each bin is $0.2$ in [C/Fe] and $\log_{10}(P/$days) respectively. The y-axis indicates the expected number of CEMP stars in each bin. The plot is normalised such that the area under the graph is the same and BHL model peaks at 1.}
   \label{fig:obs}
   \end{figure}

The effect of WRLOF on a population of binary stars is to widen towards longer periods and lower secondary masses the range of initial separations and masses of systems that become CEMP stars. As a result we calculate a CEMP/VMP ratio of $2.63\%-4.06\%$, an increase by a factor $1.2-1.8$ compared to prior studies \cite{Izzard09}, where the range of variations is due to the assumptions made on the amount of angular momentum carried away by the wind lost and on the dependence of the wind accretion efficiency on the mass ratio of the two stars. The CEMP fraction is still small compared to the observed fraction of CEMP stars.
In Fig. \ref{fig:obs} (left panel) we show the distribution of the carbon abundance calculated with the BHL (blue) and the WRLOF (red) prescriptions compared with the observations (histograms with Poisson errors). Our observed sample is based mainly on the SAGA observational database and contains all data on CEMP stars currently available in the literature \cite{Frebel2006, Lucatello05b, Suda2011}. The WRLOF model shows qualitatively the same trend as the data, but peaks at a carbon abundance higher than the observations. 
This may be an indication that in our model the accretion efficiency is too high. The transferred material is mixed in the envelope of the secondary star, composed mainly of hydrogen. A star of $0.4\,\Msun$ must accrete at least other $0.4\,\Msun$ from the donor to be selected as a CEMP star and this material is only weakly diluted in its envelope. Hence secondary stars with low initial mass become strongly carbon-enriched stars. If we reduce the accretion efficiency of the mass transfer very low-mass stars do not increase their mass enough to be selected as CEMP stars and this results in a shift of the distribution towards lower values of $[\C/\Fe]$.
In Fig. \ref{fig:obs} we compare the final period distributions predicted by our models with the observations. Even though many CEMP stars are known to be in binaries, only a few have known orbital periods because long orbital periods are difficult to measure. Our model based on the WRLOF prescription predicts most CEMP stars at orbital periods around $2000$ days, more than one order of magnitude less than the model based on the BHL prescription. Both our models have difficulty reproducing the shortest-period CEMP stars.

\section{Summary}
We include for the first time the WRLOF process of mass transfer in a binary population synthesis model and we study the effects on a population of CEMP stars. Our WRLOF model predicts an increase of the CEMP/VMP ratio up to $4.06\%$, a factor $1.8$ higher than the predictions with the BHL prescription. However, this ratio is still low compared to the observed CEMP fraction. The WRLOF period distribution peaks at shorter periods compared to the canonical BHL model but the two shortest-period CEMP stars are not reproduced. The carbon distribution predicted by our WRLOF model is qualitatively similar to the observations but it peaks at a higher value of $[\C/\Fe]$. 

\acknowledgments{SdM acknowledges NASA Hubble Fellowship grant HST-HF-51270.01-A awarded by STScI, operated by AURA for NASA, contract NAS 5-26555. CA acknowledges LKBF and the LOC of NIC-School and NICXII for travel support.}

\end{document}